# STIMULATED GENERATION OF MAGNETRONS POWERED BELOW THE SELF-EXCITATION THRESHOLD VOLTAGE*

G. Kazakevich[#], R. Johnson, Muons, Inc., Batavia, IL 60510, USA
T. Khabiboulline, V. Lebedev, G. Romanov, V. Yakovlev, Fermilab, Batavia, IL 60510, USA

*Abstract*
Modern CW or pulsed superconducting accelerators of megawatt beams require efficient RF sources controllable in phase and power. It is desirable to have an individual RF power source with power up to hundreds of kW for each Superconducting RF (SRF) cavity. For pulsed accelerators the pulse duration is in the millisecond range. The efficiency of the traditional RF sources (klystrons, IOTs, solid-state amplifiers) in comparison to magnetrons is lower and the cost of a unit of RF power is significantly higher. Typically, the cost of RF sources and their operation is a significant part of the total project cost and operation. The magnetron-based RF sources with a cost of power unit of 1-3 dollars per Watt would significantly reduce the capital and operation costs in comparison with the traditional RF sources. This arouses interest in magnetron RF sources for use in modern accelerators. A recently developed kinetic model describing the principle of magnetron operation and subsequent experiments resulted in an innovative technique producing the "stimulated" generation of magnetrons powered below the self-excitation threshold voltage. The magnetron operation in this regime is stable, controllable in phase and power, and provides higher efficiency than other types of RF power sources. It allows operation in CW and pulsed modes (at large duty factor). For pulsed operation this technique does not require pulse modulators to form RF pulses. It also looks like as a promising opportunity to extend magnetron life time. The developed technique, its experimental verification and a brief explanation of the kinetic model substantiating the technique are presented and discussed in this article.

## INTRODUCTION

Magnetrons are presently used in normal conducting compact accelerators as efficient and inexpensive RF self-exciting generators. For superconducting accelerators, the magnetron generators must provide phase and power control with the rates necessary to stabilize the accelerating voltage in SRF cavities [1]. The use of a phase-modulated injected signal to control the phase of the magnetron was first described in [2]. Methods for control of the magnetron phase and power with the required rates by injecting the resonant (injection-locking) RF signals have been developed recently [3-5]. These methods would provide stabilization of accelerating voltage in a cavity both in phase and amplitude. The phase control is provided via controlling the resonant injected phase-modulated RF signal. The amplitude control is performed via the phase control using vector methods [3, 4] or via the magnetron current control [5]. The last method provides higher transmitter efficiency at a required range of amplitude control with a bandwidth (presently feasible) about of 10 kHz.

For evaluation of properties of the RF-driven magnetrons and substantiation of the innovative technique an analytical kinetic model [6] has been developed. The model considers the basic principle of magnetron generation - the resonant interaction of the flow of the phase-grouped Larmor electrons with a synchronous wave. This interaction results in an energy exchange between the wave and the electrons. The developed model enables evaluation of the necessary and sufficient conditions for the coherent generation of the magnetron and substantiates coherent generation of the tube below the self-excitation threshold voltage unlike traditionally used model of the "Reflecting amplifier" [7], which considers only forced oscillation.

The model predicts and substantiates start-up and stop of the coherent generation "stimulated" by the resonant signal injected into a magnetron, if the magnetron is powered below the self-excitation threshold voltage. When the magnetron is powered in such a way, it spends the injected resonant signal energy for the phase grouping required for the coherent generation. As follows from the kinetic model, a generation below the self-excitation threshold enables maximizing the efficiency of the magnetrons in a wide range of power control, a reduction of the magnetron noise by the injected signal and looks to be a promising way to extend the magnetron lifetime. A brief description of the developed model and advantages of the "stimulated" pulsed coherent generation of magnetrons powered below the self-excitation threshold voltage for modern megawatt superconducting accelerators are discussed.

## THE RESONANT INTERACTION OF ELECTRONS WITH THE SYNCHRONOUS WAVE IN MAGNETRONS

One considers the analytical model, based on the charge drift approximation [8], for magnetrons driven by a resonant RF signal. We discuss a conventional CW $N$-cavities magnetron ($N$ is an even number) with a constant uniform magnetic field $H$ which is above the critical magnetic field. The magnetron operates in the $\pi$-mode, i.e. with the RF electric field shifted by $\pi$ between neighbouring cavity gaps. In this simplified model we neglect the impact of space charge on the motion of electrons.

___________________
* Supported by Fermi Research Alliance, LLC under Contract No. De-AC02- 07CH11359 with the United States DOE in collaboration with Muons, Inc.
[#] gkazakevitch@yahoo.com; grigory@muonsinc.com

In a steady-state there is a standing wave with an RF electromagnetic field induced by the injected resonant RF signal and the magnetron current in the magnetron interaction space. The wave may be presented as a sum of two travelling waves rotating with angular velocity of $\Omega=2\cdot\omega/N=\omega/n$ in opposite directions. The wavelength, $2\pi c/\omega$, is much larger than the Larmor radius $r_L$. Thus, in the drift approximation one can consider the interaction of the rotating waves with an electron in magnetic field as an interaction with the electron charge located in the center of the Larmor orbit. We consider generation of the magnetron as an energy exchange between the RF field and the drifting charge, or in other words, between the rotating waves and the drifting charge. Note that the wave rotating in the opposite direction to the drift does not participate in energy exchange with a drifting charge [9]. The condition of the $\pi$-mode generation implies that one needs to consider interaction of the drifting charge with the wave rotating in the same direction. Moreover, the average azimuthal velocity of the drifting charge has to be close to the azimuthal velocity of the rotating wave. Such a wave is named the synchronous wave. Below it will be shown that the resonant interaction of drifting charges with a synchronous wave of sufficient amplitude causes a coherent generation of the magnetron.

Thus, we consider motion of the charge drifting in the crossed constant fields in the presence of the azimuthal and radial electric fields of the synchronous wave [6].

Due to small size of the interaction region the synchronous wave radial and azimuthal fields can be approximated by the resonant harmonic of a scalar potential [6] satisfying the Laplace equation. Then, in the rotating frame of the synchronous wave the drifting charge motion is described by the following equations [10, 6]:

$$\begin{cases} \dot{r} = \omega \dfrac{r_S^2}{r} \varepsilon \phi_1(r) \cos(n\varphi_S) \\ n\dot{\varphi}_S = -\omega \dfrac{r_S^2}{r} \left( \dfrac{d\phi_0}{dr} + \varepsilon \dfrac{d\phi_1}{dr} \sin(n\varphi_S) \right) \end{cases}.$$

Here: $r_S = \sqrt{-ncU/(\omega H \ln(r_2/r_1))}$ is the synchronous radius on which the azimuthal velocities of the drifting charge and the synchronous wave coincide; $r_1$ and $r_2$ are the magnetron cathode and anode radii, respectively; parameter $\varepsilon$, is the ratio of the radial electric field of the synchronous wave to the static electric field, both taken at the cathode $\left(\varepsilon = \widetilde{E}_n(r_1)/E_r(r_1) = \widetilde{E}_n(r_1)\cdot r_1 \cdot \ln(r_2/r_1)/U\right)$; $U$ is the magnetron cathode voltage; $n\varphi_S=n\varphi+\omega t$, $\varphi$ is the phase of the emitted charge in the rest system; the terms $\phi_0$ and $\phi_1$ are:

$$\phi_0(r) = \ln\dfrac{r}{r_1} - \dfrac{1}{2}\left(\dfrac{r}{r_S}\right)^2 \text{ and } \phi_1(r) = \dfrac{1}{2n}\left[\left(\dfrac{r}{r_1}\right)^n - \left(\dfrac{r_1}{r}\right)^n\right].$$

From the above equations one can see that the drift of charges towards the magnetron anode is possible only in "spokes" at $0 < \varepsilon \leq 1$ in the phase interval $(-\pi/2 \leq n\varphi_S \leq \pi/2)$ with a period of $2\pi$. The lower equation shows the phase grouping of the drifting charges into "spokes" which carry drifting charges towards the magnetron anode. The phase grouping of the Larmor orbits into "spokes" is provided by the azimuthal electric field of the synchronous wave.

The equations were integrated for parameters of typical commercial magnetron [6]. This resulted in charge trajectories at $r \geq r_1+r_L$ for various magnitudes $\varepsilon$ of the RF field in the synchronous wave and during the time interval of the drift of $2\leq\tau\leq10$ cyclotron periods allowing coherent contribution to the synchronous wave. We assume that $\tau<2\cdot2\pi/\omega$ is insufficient to form a "spoke". Actually, at $\tau>10\cdot2\pi/\omega$ the contribution to the coherent generation is negligible [8]. Trajectories plotted in Fig. 1 [6] show motion of the charges emitted in various phases $n\varphi_s$ and drifting towards the anode.

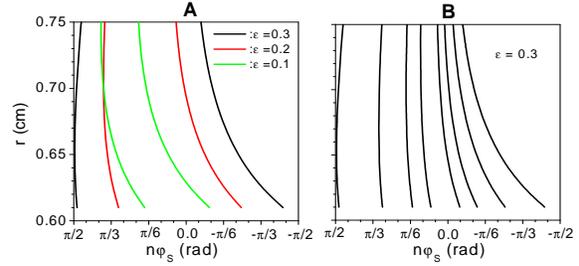

Fig. 1: A - boundaries of the "spoke" at various $\varepsilon$. B - trajectories of the charges drifting towards the magnetron anode at $\varepsilon =0.3$. The trajectories demonstrate a phase grouping of the emitted charges.

The solution of the drift equations for the typical magnetron model shows that the value $\varepsilon\approx0.3$ provides the phase grouping almost all of the emitted electrons as it is required for optimal magnetron generation. Such a value of $\varepsilon$ minimizes loss of the charge delivered to the anode [5] or in other words this reduces the charge returning back to the cathode. This phenomenon is expected to extend the magnetron lifetime.

Since in the steady-state the fields of the synchronous wave induced by the injected resonant signal and the magnetron current are in phase, an increase of $\varepsilon$ via the injected signal improves the phase grouping at insufficient $\varepsilon$. This increases the charge delivered in a "spoke" towards the anode and may allow the magnetron generation start-up when the cathode voltage is below the magnetron self-excitation threshold voltage [5, 6].

The magnetron coherent generation is provided by so-called resonant energy exchange between the phase-grouped charges and the rotating synchronous wave. In the rotating frame of the slow synchronous wave its azimuthal electric field can be considered as stationary with zero phase velocity $\Omega/n =0$. Here $n=N/2$ is the rotating field wave number. The synchronous wave electric field coupled with the resonant mode of the magnetron oscillation acts as a stationary field on the charge drifting in the "spoke". This causes the resonant energy exchange between the synchronous wave and the drifting charge. If the azimuthal velocity of the drifting charge is greater than the azimuthal velocity of the synchronous wave the charge is decelerated [11] thus increasing the amplitude of the synchronous wave and the RF field in the entire magnetron system. Otherwise, the electric field of the wave accelerates the charge increasing its azimuthal drift velocity. This reduces the wave energy and its amplitude. The resonant energy exchange of a drifting charge with a synchronous wave is the basic principle of the magnetron generation.

The increase or decrease of the self-consistent electric field of the synchronous wave resulting from the resonant energy exchange with a charge drifting in a "spoke" can be determined by the difference in the azimuthal velocities of the drifting charge and the synchronous wave: $\Delta n\dot{\varphi}_S(n\varphi_S,r)=n\dot{\varphi}_S(n\varphi_S,r)-\omega r/(n\cdot c)$ calculated for the drift, here $r$ is the current radius of the charge trajectory [6]. Note that at the synchronous radius $r_S$, the azimuthal velocity of the synchronous wave for the typical magnetron model is $\approx 0.074\cdot c$. The energy decrement or the energy increment in the synchronous wave may be determined by the sign of the $\Delta v_{AZ}(n\varphi_S, r)$ quantity, and one can estimate the value of $\varepsilon$ necessary and sufficient for the coherent generation of the magnetron by integration of $\Delta v_{AZ}(n\varphi_S, r)$ over the entire phase interval admissible for a "spoke" along the charges trajectories. Positive values of the integrals indicate coherent generation of the magnetron, while the negative values indicate damping of the synchronous wave and stop of generation.

Calculated values of the integrands $\Delta v_{AZ}(n\varphi_S, r)$ in dependence on phase of the charge emission and $\varepsilon$ values for a typical magnetron model are plotted in Fig. 2, [6].

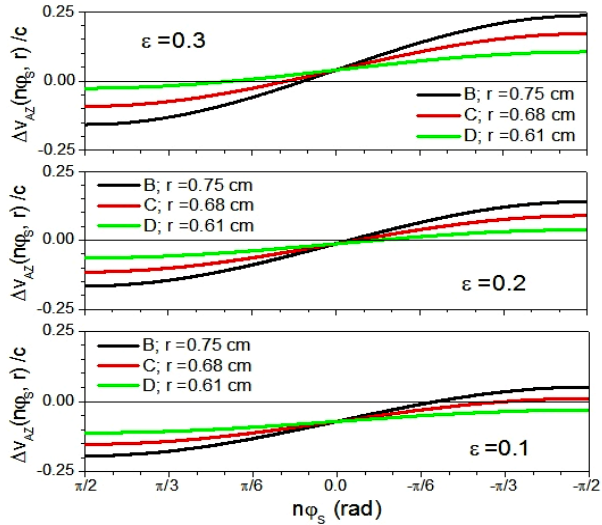

Fig. 2: Difference of the azimuthal velocities of the drifting charge and the synchronous wave in units of $c$ at various $r$, and $\varepsilon$ vs. $n\varphi_S$.

The plots indicate that at $\varepsilon =0.3$ for all radii of their trajectories the charges provide increment to the self-consistent field since the integrals are positive. At $\varepsilon \leq 0.2$ for all radii dominates a decrement causing damping of the self-consistent field. The drift equations solution and plots shown in Fig. 2 indicate that the main contribution to the self-consistent RF field provide the charges emitted in the phase interval of $h\varphi_S$: $(-\pi/2, -\pi/3)$. Note that the distribution function of the azimuthal velocities of the drifting charges at $\varepsilon =0.3$ is negative, while at $\varepsilon \leq 0.2$ it is positive.

As in the case of a small injected signal [12], the impact of a sufficiently large (about of -10 dB) injected signal on the output power of the magnetron is insignificant as demonstrated below.

The self-consistency of the electric field of the synchronous wave improves the phase grouping thus resulting in a further energy increment into the synchronous wave until saturation of the power in the magnetron is reached. The power is determined by the magnetron cathode voltage at a given power $P_{Lock}$ of the injected resonant signal. The anode losses in the modern high-power CW magnetrons are low. That means that the charges reaching the anode almost entirely transfer their energy to the synchronous wave. This results in an efficient coherent RF generation in magnetrons with a quite short build-up time. Note that when the injected signal is switched OFF the generation may be stopped due to the synchronous wave damping. Durations of the both processes are set by a loaded Q-factor of the magnetron RF system, which is small for typical magnetron.

Evaluations proved by experimental data showed that the injected resonant signal about of -10 dB of the magnetron nominal power provides stable coherent generation of magnetrons at a current $\leq 1/3$ of the minimum current required for operation without the injected signal. For such a low current the magnetron must be powered with the voltage below its self-excitation threshold. This allows a power control in magnetrons in the range about of 10 dB at highest efficiency, varying the magnetron current in an extended range [5].

A novel technique enables start and stop of the magnetron coherent generation without a modulation of the cathode voltage, when the magnetron is powered below the self-excitation threshold. The technique uses the synchronous wave start-up and damping when the RF injected signal is switched ON-OFF.

Experimental verification of this technique attractive for utilization in pulsed and CW superconducting accelerators with megawatts beams is presented below.

## A NOVEL TECHNIQUE OF THE MAGNETRON PULSED GENERATION

The proof of the principle of this technique was demonstrated with a CW, 2.45 GHz microwave oven magnetron type 2M219G with nominal output power of 945 W and the measured magnetron self-excitation threshold voltage of 3.69 kV. The magnetron was powered by a pulsed High Voltage (HV) source using a partial discharge of the storage capacitor, Fig. 3, [13], providing pulse duration of $\approx 5$ ms.

The HV source provided negligibly small ripple. This enables to avoid the magnetron start-up caused by the ripple. At the magnetron RF power $\approx 1$ kW and pulse duration of 5 ms a discharge of the storage capacitor causes a decrease of the magnetron cathode voltage of ≈ 0.4%. The pulsed HV source was powered by a charging Glassman 10 kV, 100 mA switching power supply allowing the voltage control.

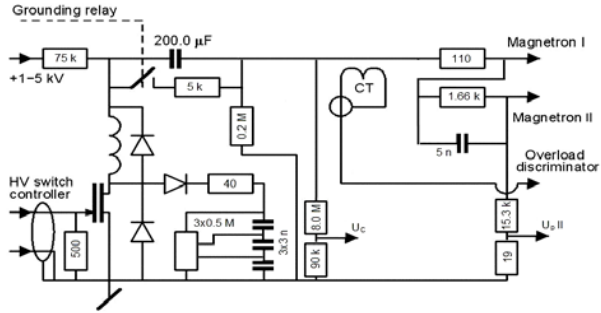

Fig. 3: Schematic diagram of the pulsed HV power supply with Behlke MOSFET IGBT 10 kV/800 A switch.

Pulsed "stimulated" generation of the magnetron was realized by the injection of a pulsed locking signal into the magnetron RF system. This has been studied with the setup shown in Fig. 4. The CW signal of an HP-8341A generator was converted to RF pulses by a mixer (ZEM-4300MH from Mini-Circuits) controlled by a pulsed generator (type 100A). Then the RF pulses were amplified by solid-state and TWT amplifiers which provided the pulsed RF signal with a power up to 160 W driving the magnetron.

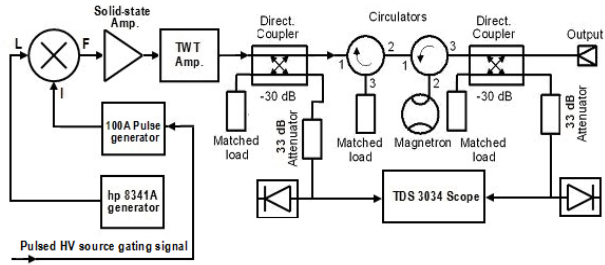

Fig.4: Setup to study the ON-OFF switch control of the magnetron driven by pulsed resonant injected signal.

Fig. 5 shows 4 kHz trains of 147 μs RF pulses for the injected and the magnetron output signals, as well as the magnetron current and the HV pulse of 5.1 ms duration. One can see the magnetron switching ON-OFF due to the injected resonant signal without pulsed modulation of the cathode voltage.

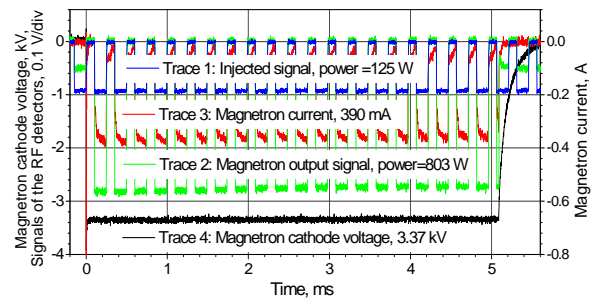

Fig. 5: 4 kHz trains of 147 μs pulses (duty factor of ≈59%). Traces 1 and 2 - the resonant injected and the magnetron output RF signals with powers of 125 W and 803 W, respectively; trace 3 - the magnetron pulsed current; trace 4 - the magnetron cathode voltage.

In the first measurements we did not synchronize the injected resonant signals with our pulsed HV source. This allowed studying the magnetron ON-OFF control at the fronts of high voltage pulses. Pulse shapes and power levels of the injected signal and the magnetron output signal were measured by the RF detectors with Schottky zero-bias diodes calibrated with inaccuracy less than ±0.5%. The magnetron pulse current was measured by a current transducer (type LA 55-P) with the circuit integration time of about 50 μs.

The magnetron current is decreased by 4.4±0.5% during the 5.1 ms HV pulse. The decrease is caused by a small discharge (~0.4%) of the HV source storage capacitor. This results in the measured decrease of the magnetron output power (≈ 4.9%). A part of power of the injected signal (<40% of the locking power $P_{Lock}$) goes to the magnetron output. It is clearly seen when the magnetron cathode voltage is OFF.

## PROPERTIES OF THE MAGNETRON GENERATING BELOW THE SELF-EXCITATION THRESHOLD

Unlike the start-up and stable operation of magnetrons driven by a sufficient resonant signal and powered below and above the self-excitation threshold voltage, operation of the magnetron in the mode of "stimulated" generation imposes an additional limitation on the cathode voltage to stop the generation when the injected signal is switched OFF. Measured maximum and minimum of the magnetron cathode voltage and current allowing the "stimulated" pulsed generation of the magnetron 2M219G at various power of the injected signal, $P_{Lock}$ are plotted in Fig. 6. The solid lines in the plot show the ranges of the magnetron cathode voltage (and the magnetron current) allowing stable operation in the "stimulated" generation mode. The injected signal duration at the measurements was 2.2 ms at the magnetron cathode voltage duration of 5.1 ms.

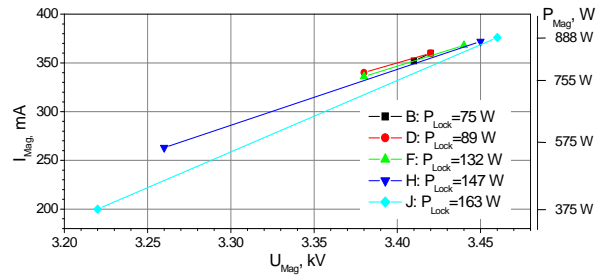

Fig. 6: The ranges of the 2M219G magnetron cathode voltage and the magnetron current in the "stimulated" generation mode at various power levels of the injected resonant signal $P_{Lock}$. The right scale shows measured RF power of the magnetron vs. the magnetron current, $I_{Mag}$.

The measurements show that with "stimulated" generation the magnetron can provide almost nominal output power. at a lower cathode voltage. This is due to the improved phase grouping and reduced losses of drifting charges with sufficient amplitude of the synchronous wave.

Fig. 6 indicates possibility of power control of the magnetron by regulation of the cathode voltage in an admissible interval depending on $P_{Lock}$. Note that the relative variation of the magnetron output RF power, $\Delta P_{RF}/P_{RF}$, requires much smaller variation of the magnetron cathode voltage $\Delta U_{Mag}/U_{Mag}$:

$$\Delta P_{RF}/P_{RF} \approx (Z_S/Z_D) \cdot \Delta U_{Mag}/U_{Mag}.$$

Here $Z_S$ and $Z_D$ are the static and dynamic magnetron impedances; typically $Z_S \geq 10\, Z_D$.

The measurements indicate that an increase of the injected signal increases the difference between maximum and minimum power of the RF pulse generated by magnetron. The maximum and minimum powers are determined mainly by the maximum and minimum cathode voltage.

Fig. 7 showing the maximum and minimum powers of the magnetron operating in the "stimulated" generation mode vs. the $P_{Lock}$ value demonstrates that the injected RF signal of -8 dB allows a regulation of the magnetron power in the range of ≈ 3 dB. For the magnetron power regulation in the range of 7 dB the injected signal with $P_{Lock} \approx 170$ W (-7.4 dB) is required.

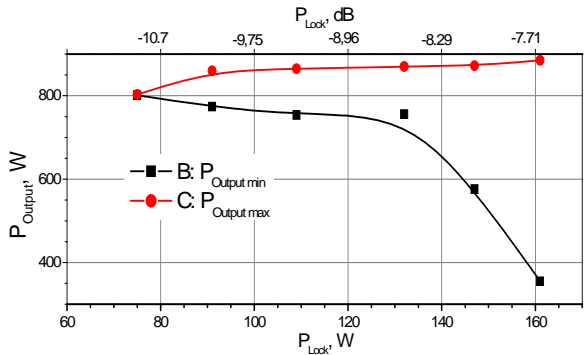

Fig. 7: Measured maximum and minimum output power levels of the 2M219G magnetron operating in the "stimulated" mode vs. $P_{Lock}$.

Using a two-stage magnetron RF source [3, 6], composed of small and high-power magnetrons (both tubes operate in a "stimulated" generation mode) it is possible to reduce power of the resonant signal controlling the RF source by ≈10 dB. This will reduce the capital cost of the driver module. In this case a powerful magnetron can provide highly efficient power control in the middle frequency band and broadband phase control by a resonant driving signal with power of about -17 dB from the nominal power of powerful magnetron.

The magnetron with cathode voltage above the ranges shown in Fig. 6, but still below self-excitation threshold, continues generation even after the injected signal which is switched OFF, Fig. 8. The continuation of the generation in this case in the absence of the injected signal can be explained by the finite damping time of the synchronous wave.

Shown in Fig. 8 the noise in traces of the magnetron current and the RF output power is appeared because of loss of coherency when the locking signal is switched OFF, but the generation is continued at the cathode voltage less than the self-excitation threshold [6].

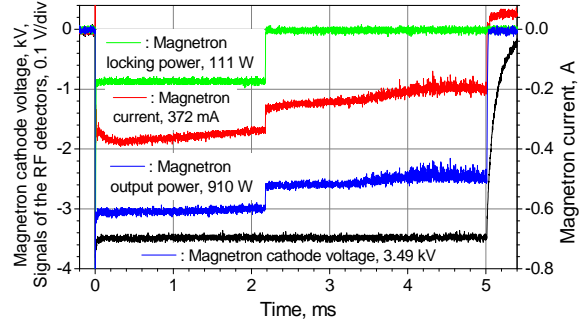

Fig. 8: Continuation of the magnetron generation at a magnetron voltage higher than it is required for the "stimulated" generation at the given $P_{Lock}$ after the injected signal is switched OFF.

Measured with better time resolution the traces of a 20 kHz train of 13 µs long RF signals injected into the magnetron and the output signals, are shown in Fig. 9.

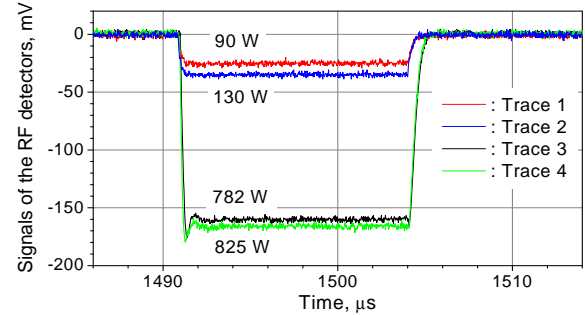

Fig. 9: 13 µs pulses of the measured 20 kHz trains when the magnetron operates in the "stimulated" generation mode. Traces 3 and 4 are the magnetron output RF signals in dependence on the power of driving signals shown in traces 1 and 2, respectively.

All traces in Fig. 9 demonstrate relatively short rise and fall times when the magnetron is switched ON or OFF by the injected resonant signal. These times do not exceed 200 ns that roughly correspond to ~ 200 cyclotron periods.

These measurements were performed at a magnetron cathode voltage of 3.41±0.01 kV. This voltage is sufficient for the "stimulated" generation at the injected signals of 90 and 130 W. When the magnetron is switched OFF, ≈40 W of the injected signal (a few percent of the magnetron nominal power) goes to the magnetron output. Almost equal differences in power of the magnetron injected and output signals indicate quite weak impact of the injected signal level on the magnetron output power even at a large injected signal. This resembles a negligible impact of the low-power injected signal considered above.

Large noise in the magnetron output when the magnetron voltage is close to the excitation threshold is known. Noise may be caused by relaxation oscillations in the magnetron-HV power supply system [6], due to the loss of coherence with insufficient $\varepsilon$, (as it demonstrates Fig. 8) or/and due to the ripple of the HV power supply. The operation of the magnetron in the mode of "stimulated" generation was tested for the presence of such noise. Figs. 9 and 10 clearly show an absence of noise when the generation of the magnetron is switched ON-OFF using the injected resonant signal, even if the voltage of the magnetron cathode is slightly lower than that required for "stimulated" generation with this injected signal. The decrease in magnetron voltage which stopped its generation, caused by some discharge of the storage capacitor of the high-voltage power supply was used for the measurements shown in Fig. 10. An absence of noise indicates an absence of the necessary and sufficient conditions for generation when the injected resonant signal is switched OFF and the magnetron cathode voltage corresponds to the limitation of the "stimulated" generation mode being below the self-excitation threshold.

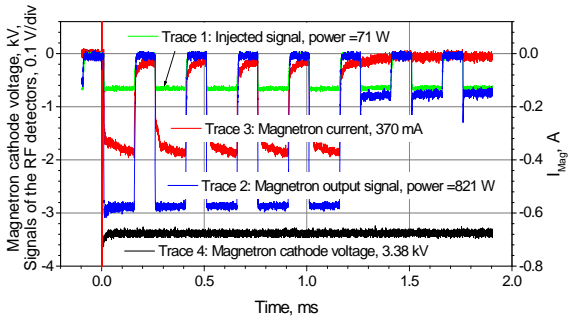

Fig. 10: 4 kHz trains of 147 μs pulses: traces 1 and 2 - the resonant injected and the magnetron output RF signals with powers of 71 W and 821 W, respectively; trace 3 - the magnetron pulsed current (right scale); trace 4 - the magnetron cathode voltage.

When the magnetron cathode voltage become a bit less than it is required for the "stimulated" generation the insufficient phase grouping of Larmor electrons stops the magnetron generation, and on the output corresponds to a signal with power approximately equal to the injected one, Fig. 10.

When voltage of the magnetron cathode is OFF, Fig. 5, the phase grouping does not exist, and the output signal power is about of 30% of the injected power.

Comparison of the both cases indicates that almost entire power of the injected signal is spent for the phase grouping when the cathode voltage is a bit more than it is required for the magnetron start up. Stop of the magnetron generation releases the power spent for phase grouping sending it to the magnetron output.

The magnetron operation is characterized by its conversion efficiency, $\eta$, which expresses the transformation of consumed DC power to the generated RF power, $P_{RF}$: $\eta \approx P_{RF}/(U_{Mag} \cdot I_{Mag} + P_{Lock})$. Here $U_{Mag}$ and $I_{Mag}$ are the measured values of the magnetron voltage and magnetron current, respectively. Typically, the magnetron filament power consumption does not exceed a few percent and one can neglect this contribution. The measured conversion efficiency of the 2M219G magnetron for various powers of the injected signal in accordance with minimum and maximum output power admissible for the "stimulated" generation mode is shown in Fig. 11.

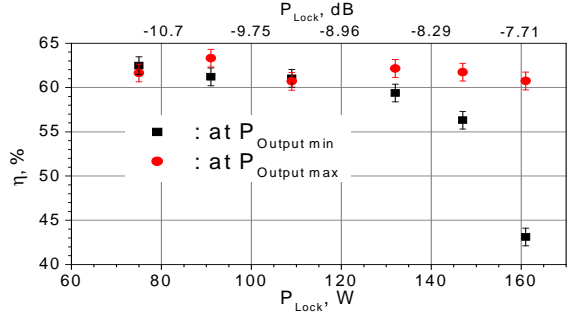

Fig. 11: Dependence of conversion efficiency of the 2M219G magnetron on power of injected signal $P_{Lock}$.

The measured conversion efficiency of magnetron operating in self-excitation ("free run") mode ($U_{Mag} \approx 3.69$ kV, $P_{Lock} = 0$), is $\approx 54\%$. For the output power of $\approx 900$ W this conversion efficiency corresponds to the anode and cathode losses in the magnetron about of 414 W, while they do not exceed 338 W in "stimulated" generation mode. The difference about of 76 W exceeds power of the filament heating of the magnetron cathode more than 2 times. We can assume that the larger part of this power relates to the cathode losses resulting in overheating the cathode emitting surface.

The measurements shown in Fig. 11 demonstrate that the magnetron operating in "stimulated" mode provides higher conversion efficiency. As it was noted above, the developed analytical model predicts the higher efficiency as a result of the improved phase grouping for magnetrons operating at a sufficient resonant injected signal. The improved phase grouping reduces loss of charge carried by "spokes" to the magnetron anode [5], reducing thus the electron back-stream overheating the emitting surface.

Prolongation of the injected resonant signal up to duration of operation of the cathode voltage power supply did not demonstrate any violations of the "stimulated" pulsed generation of the magnetron, Fig. 12.

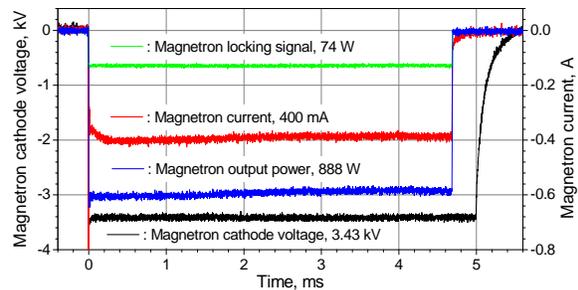

Fig. 12: Operation of the magnetron in the "stimulated" mode of pulsed generation with pulse duration of 5 ms.

Small decreases of the magnetron current and power along the pulse are caused by a discharge of storage capacitor of HV power supply as was noted above.

## ON THE LIFE EXPECTANCY OF HIGH-POWER MAGNETRONS FOR SUPER-CONDUCTING ACCELERATORS

Utilization of the high-power CW magnetron as RF sources for GeV-scale accelerators with megawatts beams was considered in a number of works, e.g., [14]. But, typically the industrial magnetrons are designed as RF sources for heating, so the lifetime of tubes is not the first priority as it is required for accelerators.

The high-power industrial CW magnetrons use the cathodes made of pure tungsten. The emission properties of the tungsten cathodes are not deteriorated much by electron and ion bombardment, but the latter causes the sputtering of the cathode. The sputtering leads to sparks and discharges in the space of interaction of magnetron reducing its lifetime.

Typical current density emitted by the tungsten cathodes is about of 0.3 A/cm$^2$ in the industrial magnetrons [15] which requires the cathode temperature of about 2500 K. Operation of the cathode at such high temperature and the cathode additional overheating caused by electron back-stream worsen the vacuum in the cathode vicinity approximately by an order of magnitude. In the cold magnetron the vacuum pressure is typically about $10^{-6}$ torr, while during tube operation the average vacuum pressure in the space of interaction is $\sim 5 \cdot 10^{-6}$ torr. For 100 kW CW magnetrons the cathode voltage is ≈20 kV. Taking into account the efficiency of RF generation it can be assumed that the average electron energy in the interaction space is $\sim 3$ keV. The electrons ionize the residual gas. This results in appearance of number of ions accelerated by the static electric field towards the cathode. Assuming, that the residual gas is nitrogen with the cross section of single ionization $\sim 3 \cdot 10^{-17}$ cm$^2$ [16], one can estimate the flux of ions hitting the cathode with the energy determined by the static electric field.

For the commercial 100 kW, 920 MHz magnetron [15] the probability to create an ion on the length of Larmor orbit is $3.8 \cdot 10^{-5}$. For the emitted current of $\sim 6$ A the ion flux is $\sim 5 \cdot 10^{14}$ ions/s, i.e., the flux density of ions hitting the cathode is about of $7.2 \cdot 10^{13}$ ions/cm$^2 \cdot$s with the average ion energy of $\sim 10$ keV. This corresponds to the power density of the ion flux $\sim 0.1$ W/cm$^2 \cdot$s.

Rough estimate for the sputtering yield from tungsten caused by 10 keV nitrogen ions is $\sim 0.03$ atom/ion [17]. For 1000 h of operation loss of tungsten by the cathode is $\sim 0.05$ g or $\sim 25$ μm thick layer with area ≈19.3 cm$^2$. The sputtered material of the cathode covers the magnetron interior including the coupling windows of the resonant cavities. This causes sparks and discharges in the magnetron. Note also that the number of sputtered atoms can be even higher due to increased sputtering at high temperature and the cathode evaporation [17].

The above estimates indicate that the main reason limiting magnetron lifetime is the cathode sputtering caused by ionization of the residual gas. This phenomenon is inherent to operation of magnetrons. A decrease of the vacuum pressure in magnetrons to the value of $10^{-8}$ torr or less, e.g., by the built-in vacuum pump will decrease the cathode sputtering thus increasing the magnetron longevity.

A lower cathode voltage at the "stimulated" generation mode somewhat reduces probability of discharges and sparks in the space of interaction which are caused by the cathode sputtering. Also a reduction of the electron back-stream in the "stimulated" generation mode somewhat reduces the cathode temperature.

Thus, it is expected that improved vacuum in the tube by approximately two orders of magnitude along with the operation of magnetron in the "stimulated" generation mode will decrease the sputtering and its impact. That promises to increase lifetime of a high-power magnetrons and make them suitable RF sources for GeV-scale accelerators with megawatts beams.

## SUMMARY

We have developed an innovative technique allowing a pulsed RF generation of a magnetron without modulation of the cathode voltage, while the tube is powered below the self-excitation threshold. A pulsed resonant injected signal with power ≥ -11 dB of the magnetron nominal power is required for such operation. The technique was substantiated by the developed kinetic model of the magnetron operation representing the principle of RF coherent generation of the tube. Experiments with a 2.45 GHz magnetron proved the developed technique. The CW magnetron operating in the "stimulated" mode has efficiency higher than for its operation in a "free run" or driven by small (≤ -20 dB) injected signal. The latter cases relate to the traditional regimes when the tube operates above the self-excitation threshold. The magnetrons powered below the self-excitation threshold and driven by the sufficient resonant signal provide higher conversion efficiency, significantly lower (by ≈ 20 dB/Hz) spectral power density of noise and significantly wider (up to hundreds of kHz) the phase control bandwidth.

The developed innovative mode of magnetron operation demonstrates properties attractive for superconducting CW and pulsed accelerators with megawatts beams.

Performed estimates of the cathode sputtering in high-power CW magnetrons support expectations for extended lifetime of the magnetrons applicable for modern accelerator projects.

## ACKNOVLEDGEMENT